\documentclass[twocolumn,twoside,slac_two]{revtex4}
\usepackage{graphicx}
\usepackage{fancyhdr}
\newcommand\aap{{A\&A}}%
\newcommand\apss{{Ap\&SS}}%
\def\apj{{\em Astrophys. J.}}
\def\apj{{\em Astrophys. J.}}
\def\aap{{\em A\&A}}
\def\apjl{{\em Astrophys. J.Lett.}}

\begin{document}

\title{Lunar gamma ray emission seen during the first year by Fermi}

%

\author{N.Giglietto on behalf of the \emph{Fermi} Large Area Telescope Collaboration}
\affiliation{INFN Bari and Dipartimento Interateneo di Fisica- Universit\`a e Politecnico di Bari}
%

\begin{abstract}
We report the detection of the lunar  gamma-ray emission during 
the first year of Fermi-LAT observations. Such emission is produced by 
cosmic ray nuclei interacting with the  lunar surface. Thanks to the solar
 minimum conditions and the reduced effects of heliospheric modulation, the lunar  
flux was at its maximum due to the increased flux of Galactic cosmic rays hitting the lunar
 surface. Fermi-LAT instrument has a superior sensitivity, angular resolution, and observes 
the whole sky every two orbits. It is the only gamma-ray mission capable of detecting the lunar 
emission with high confidence and to monitor it over the full 24th solar cycle. We also report 
the status of a search of the  gamma-ray emission from major planets and asteroid populations 
in the ecliptic plane.

\end{abstract}

\maketitle

\thispagestyle{fancy}


\section{Introduction}

The $\gamma$-ray emission produced by  solid solar system bodies is due to the
interactions of Galactic cosmic ray nuclei (mainly protons) with their surface layers.
The main
processes involved are the production and decay of $\pi^0$s and kaons by ions,
bremsstrahlung by electrons  and Compton scattering  of the secondary photons.
 The $\gamma$-ray telescope EGRET on the \emph{Compton Gamma-Ray
Observatory} (\emph{CGRO}), operated from 1991 to 2000 and detected the 
$\gamma$-ray  emission from the
Earth \citep{Petry2005}, the Moon \citep{EGRET,OrlandoStrong2008}, and
the Sun \citep{OrlandoStrong2008}.
Nuclear  $\gamma$-ray  emission produced in the lunar surface in low-energy  
cosmic ray and secondary neutron reactions was observed by 
the Lunar Prospector \citep[][]{LunarProspector} and used in the 
analysis of the composition of the regolith. 
Although similar physical processes are involved, the $\gamma$-ray spectra
of the Earth, the Moon, and the Sun 
are very different.
The Moon is so far the only observed $\gamma$-ray 
emitting body with the solid surface. For the Sun,  
the   $\gamma$-ray emission from the disk, due to the interactions of cosmic ray
nuclei with the solar atmosphere \citep{Seckel1991,OrlandoStrong2008},
is accompanied by extended and brighter 
$\gamma$-ray emission due to the inverse Compton scattering of
Galactic cosmic ray electrons off solar photons
\citep{Moskalenko2006,OrlandoStrong2007,OrlandoStrong2008}.

Early analysis of EGRET observations of the Moon yielded the integral
flux of $F(E>100\ {\rm MeV})=(4.7\pm0.7\ {\rm syst})\times 10^{-7}$
cm$^{-2}$ s$^{-1}$ \citep{EGRET}. A later reanalysis confirmed the
detection and yielded a flux $F(E>100\ {\rm MeV})=(5.55 \pm  0.65)\times
10^{-7}$ cm$^{-2}$ s$^{-1}$ averaged over the entire mission duration
\citep{OrlandoStrong2008}. 

Calculations of interactions of cosmic rays with the lunar surface are
fairly straightforward and involve a well-measured spectrum and
composition of cosmic rays near the Earth and composition of the Moon
rock, the regolith, which was quite rigorously studied using the
samples returned by the lunar missions as well as by the remote
sensing.
The first calculation of the lunar $\gamma$-ray emission was done by Morris\cite{Morris84}
 using the cross section data and techniques
available at that time.  
Recent detailed flux calculations\citep{Igor} depends on the level of 
heliospheric modulation and is generally consistent with observations and
were done using the Geant4 framework\citep{GEANT4}. 
It was shown that the spectrum of
$\gamma$-rays from the Moon is steep with an effective cutoff
around 3--4 GeV (600 MeV for the inner part of the lunar disk).
Due to the kinematics of the collision, the secondary
particle cascade from cosmic ray particles hitting the lunar surface
at small zenith angles develops deep into the rock making it difficult
for $\gamma$-rays to get out. Therefore the lunar $\gamma$-ray emission is produced by a small
fraction of splash  particles in  the surface layer of the
moon rock. High energy $\gamma$-rays can be produced by cosmic
ray particles hitting the Moon surface with a more  tangential
trajectory; thus  only a very thin
limb contributes to the high energy emission.

A similar emission should be detected from any other solid object in the solar system.
 Therefore planets should emit $\gamma$-rays produced from pion-decays coming from the hadronic interactions by
 cosmic-rays hitting the surface or the atmosphere of these bodies.
Recent papers\cite{Igor3p, IgorHEAD} have studied interactions of cosmic rays with
 populations of small bodies in the solar system.
 In particular, gamma-ray emission from asteroids, rocks, 
and dust at the outskirts of the solar system, e.g., the Oort Cloud\cite{Moskalenko:2009tv},
 may contribute to the isotropic gamma-ray background. 
The gamma-ray emission from the Moon\cite{Igor, Igor2} can serve as a template for such studies. 

\begin{figure*}[ht]
 \centering

\includegraphics[width=130mm]{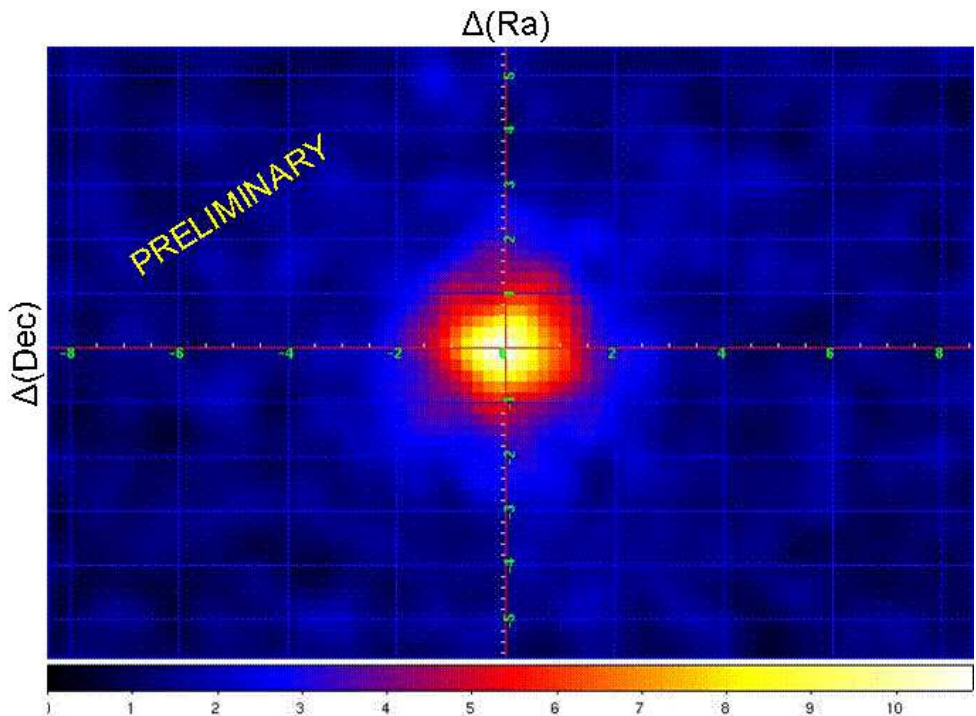}
\caption{Count map in a Moon centered frame, Right Ascension and Declination offsets in degrees respect to the Moon position in abscissa and ordinate respectively.
The offset ranges are $\pm~8^\circ$  for Right Ascension and $\pm~5^\circ$ for Declination.
  The image has been obtained by using photons with E$>$~100~MeV, a bin width of 0.2 degrees with a gaussian smoothing 2 bin radius.  
    The   colour scale is linear with the counts.}
\label{moon-map}
\end{figure*}

We report here the updated  observations of the lunar $\gamma$-ray  emission,
 previously presented\cite{vulcano,Brigida,icrc1}, and the
status of a search for emission from smallest solid objects, such as Jovian trojans\cite{icrc2}.

\section{Data selection}

The data used in this study were collected during the first year  beginning from 4 August 2008 
during the solar minimum of activity. We have
applied  a zenith cut of 105$^{\circ}$ to eliminate photons from the Earth's limb, and   
used the ''Diffuse'' class
\cite{LATpap}, corresponding to events with the highest probability of being true photons.
The analysis is performed using the standard maximum-likelihood
spectral estimator provided by the 
LAT Science Tools\footnote{{\tt gtlike} is the standard maximum-likelihood spectral estimator included in Science Tools}
package (version v9r15), which is available from  and using p6\_v3 post-launch Instrumental
Response Functions (IRFs). 
 We also used the standard Science Tools provided by the Fermi Science
Support Center, version v9r15 and IRFs (Instrumental Response Functions) version P6\_V3. 

The position of the Moon has been  computed using a JPL\cite{jpl} library interface. 
  Photon events for analysis were selected in a frame, 15 degree radius, and centered on the 
istantaneous lunar position. We used the unbinned  likelihood analysis technique typically used 
for astrophysical sources but, since the Moon moves quickly across the sky, we had to take additional 
precautions to determine the emission spectrum. 
In order to avoid strong variations of background photons,
 we excluded time intervals when the Moon was close to the galactic plane or bright sources. 
We therefore required that the  Moon was at least  30$^\circ$ from 
the galactic plane i.e. $|B_{moon}|>30^\circ$, and we remove also any time intervals in
 which any individual bright object has an angular distance less than  5$^\circ$ from the Moon.
The real difficulty in obtaining a consistent  lunar $\gamma$-ray flux is the variation of
 the diffuse galactic background and extragalactic emission 
as the Moon moves  through the  sky.
 Therefore to evaluate the diffuse background  in proper way we use the fake source method\cite{alexandreas,cassidy} 
by defining a ''fake'' moon 
that followed the lunar trajectory but  30 degrees  displaced along the true trajectory. 
   The ''fake'' Moon was therefore exposed to the same
celestial sources as the true Moon and the event  observed in the frame centered on the fake Moon 
make a good description of the diffuse background.
Moreover we verify that this background estimation is consistent with the
 background flux values measured using larger values of angular displacement along 
the lunar trajectory to define the fake source.

\begin{figure}
\includegraphics[height=90mm,width=90mm]{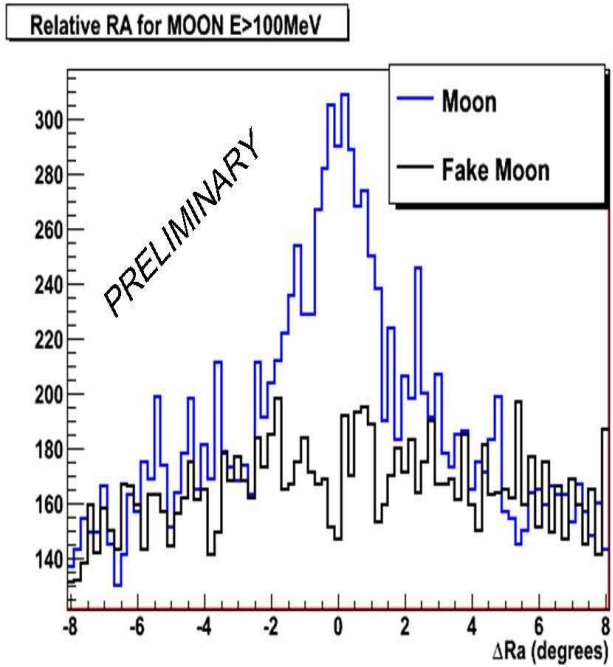}
\caption{Count map of photons having an angular distance less than 15$^\circ$ from the Moon center and with E$>$~100~MeV,
 as a function of Right Ascension offsets in degrees respect to the lunar nominal position. 
Superimposed as dashed line the fake Moon count map distribution.
 Observed data are consistent with the expected angular resolution for 100~MeV photons.}\label{Moon-ra}
\end{figure}

\begin{figure}
\includegraphics[height=90mm,width=90mm]{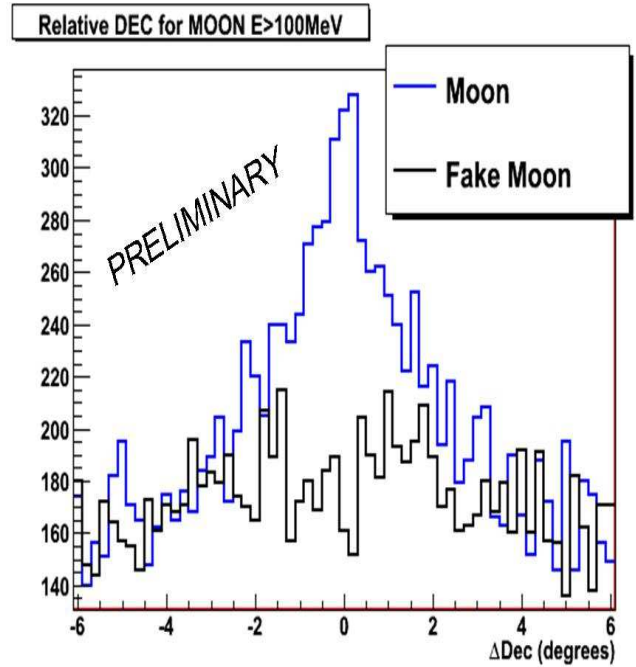}
\caption{Count map of photons having an angular distance less than 15$^\circ$ from the Moon center and with E$>$~100~MeV,
 as a function of Declination offsets in degrees respect to the lunar nominal position. Superimposed as dashed line the fake Moon 
count map distribution.
 Observed data are consistent with the expected angular resolution for 100~MeV photons.}\label{Moon-dec}
\end{figure}

\section{Lunar observation results}

Fig.~\ref{moon-map} shows the count map of photon events with E$>$~100~MeV in offsets of celestial coordinates relative to the Lunar position. Emission from the Moon is clearly visible and centered on the expected location in this relative coordinate frame.

Fig.~\ref{Moon-ra} shows the count map of the events with E$>$~100~MeV and within a 15$^\circ$ from the Moon center projected onto Right Ascension while in Fig.\ref{Moon-dec} 
the events are projected onto Declination. 
The coordinates are  offsets in degrees of celestial coordinates relative to the Moon position.
 In these figures the counts observed for the ''fake'' Moon are superimposed and demonstrate the need to carefully consider
the  background before any lunar analysis can be performed. Simulated
 data confirm that the observed shape is in agreement with a pointlike
 source and the calculated point spread funtions\footnote{see \url{http://fermi.gsfc.nasa.gov/ssc/data/analysis/documentation/}}
 of the LAT\cite{LATpap}
 for the expected lunar gamma-ray spectrum.
In a more detailed 
analysis\cite{moonpaper} of a  one-year database we will perform carefully the
 analysis of the shape of emission  to determine its
 true extent and whether we can discern evidence for the expected limb brightening.

To study the spectrum of the gamma-ray emission we have used the standard unbinned maximum-likelihood spectral estimator provided with the LAT science tools {\tt gtlike}. This preliminary analysis  was performed by fitting the 
''fake'' moon data to model the background and the Moon data sample
with either a simple power law or other functional forms like a log-parabola summed to the background model.

The fitted values obtained using a simple power
 law give a good test-statistics
 value of 7320.3 (defined as $TS=2(\log L-\log L_0)$, being $L$ and $L_0$ the likelihood values when the source is considered or not)     and indicate a power law index of $-3.13\pm0.03$.
As a result of these fits we estimate the observed flux as $F(E>100~MeV)=1.1\pm0.2\times 10^{-6}$ $cm^{-2}s^{-1}$.
The error includes
the estimation of overall systematic error of about 20$\%$ for these measurements, due essentially to the uncertainties of the detector effective area and due to  
the inefficiencies in gamma-ray detection due to pile-up effects from near coincidences with cosmic rays in the LAT detector.
A comparison of our data with models\citep{Igor} 
are in  good agreement with the  shape of 
the expected spectrum and compatible with the levels computed  of solar activity. 
The detailed analysis of the differential spectrum of the lunar emission, in particular
 below 100~MeV\cite{Igor,Igor2}, is currently in progress.

Our measurement of the lunar $\gamma$-ray flux is
about twice the average flux measured $\rm 5.5\times 10^{-7}$ cm$^{-2}$ s$^{-1}$
 previously obtained by EGRET\citep{EGRET}.
 However we know that the EGRET data subdivided 
into two subsamples, data taken during the solar maximum 1991 -- 1992 and
 those taken during period of moderate activity 1993 -- 1994,
yield to $\rm 3.5\times 10^{-7}$ cm$^{-2}$ s$^{-1}$ and 
 $7\times 10^{-7}$ cm$^{-2}$ s$^{-1}$ respectively, 
demonstrating that the flux is higher near the solar minimum.
Therefore
the {\it Fermi}-LAT observations reported in this paper, that
are taken entirely during the period of solar minimum, 
should be better compared with the EGRET subsample taken
during the solar low level of activity. The neutron monitor rate\footnote{see for example tables from \url{http://neutronm.bartol.udel.edu} or
\url{http://www.ngdc.noaa.gov/stp/SOLAR/ftpcosmicrays.html}} during
 the period of observations by Fermi was, at least, 
10\% higher 
than the ground neutron rate values reported by EGRET\citep{EGRET} at that time.
We conclude therefore that there is a reasonable agreement with previous measurements  if 
 the solar activity is taken into account.

 \begin{figure*}[th]
   \centering
  \includegraphics[width=130mm]{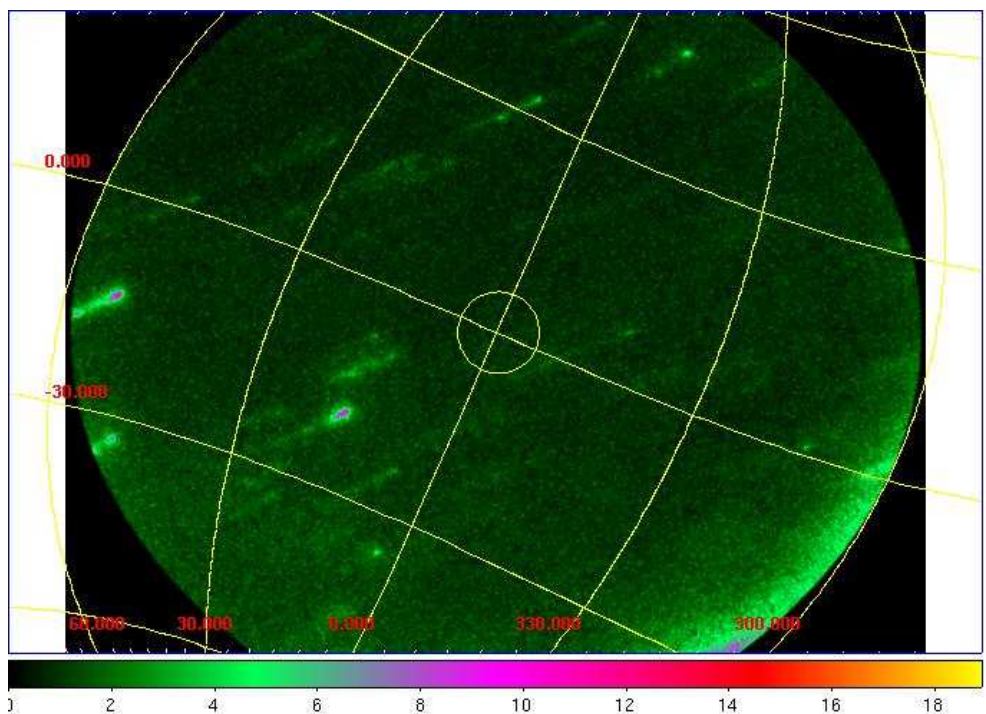}
  \caption{
Photon count map centered on Saturn position (marked with a circle) and collected during the first 10 months starting from august 2008.
 The colored vertical scale is linear and  smoothing has been applied to the image. 
The bin width used is 0.2$^\circ$. The coordinates are celestial  offset respect of Saturn position,
 the axes represent  ecliptic coordinates. 
}
 \label{saturn-map}
  
\end{figure*}

\section{Other solar system sources}
A similar strategy has been used to search for the $\gamma$-ray emission from the most interesting solar
 system objects, starting from the major planets.
We have considered the photon count maps in  the direction of major planets, 
including Jupiter and Saturn.
For any selected object we have examined the count map to look for any evident excess of counts. 
No evidence of photon excess  from the regions  centered on these planets 
 or within several degrees from these sources was found during the first 10 months of observations.
 Fig.~\ref{saturn-map} shows the count map of the events in celestial coordinates offsets
 relative to Saturn position.
The search for the emission from populations of small  
solar system bodies\cite{Moskalenko:2009tv} is currently in progress.

\section{Conclusions}
The gamma-ray emission from the Moon discovered 
by EGRET  has been confirmed by \emph{Fermi} and agrees in intensity for emission 
models that take into account the level of solar modulation. Our preliminary flux estimation
 for the lunar $\gamma$-ray emission is $F(E>100MeV)=1.1\pm0.2\times 10^{-6}\,cm^{-2}s^{-1}$ with
 a spectral index of $-3.13\pm0.03$ obtained by fitting a simple power law between 100~MeV to 1~GeV.
The lunar flux measured is higher than previous measurements but reasonably in agreement with the cosmic ray flux 
increase due the solar minimum activity in this solar cycle. 
We expect that  with a larger sample of events available after accumulating data for one year,  
\emph{Fermi} should be able to explore the other features of the lunar spectrum, e.g. the $\pi^0$ peak
and also to resolve the spatial structure of the emission from the lunar disk. 
We have started a search for the  emission from other bodies of the solar system objects too,
 in particular from Saturn and Jupiter regions. During the first 10 months we don't have evidence of $\gamma$-ray 
emission in these regions. 
The search for the emission from populations of small  
solar system bodies is currently in progress.

\bigskip 

\section*{Acknowledgements}
The \emph{Fermi} LAT Collaboration acknowledges support from a number of agencies and institutes 
for both development and the operation of the LAT as well as scientific data analysis. These include
 NASA and DOE in the United States, CEA/Irfu and IN2P3/CNRS in France, ASI and INFN in Italy, MEXT, KEK,
 and JAXA in Japan, and the K.~A.~Wallenberg Foundation, the Swedish Research Council and the National Space
 Board in Sweden. Additional support from INAF in Italy for science analysis during the operations phase is also
 gratefully acknowledged.


\end{document}